# AN EFFECTIVE PRIVACY-PRESERVING DATA CODING IN PEER-TO-PEER NETWORK

Ngoc Hong Tran[1], Cao Vien Phung [2], Binh Quoc Nguyen [1], and Leila Bahri [3]

Vietnamese-German University [1], Vietnam, Technische Universität Carolo-Wilhelmina zu Braunschweig[2], Germany, and KTH [1], Sweden

## ABSTRACT

*Coding Opportunistically (COPE) is a simple but very effective data coding mechanism in the wireless network. However, COPE leaves risks for attackers easily getting the private information saved in the packets, when they move through the network to their destination nodes. Hence in our work, a lightweight cryptographic approach, namely SCOPE, is proposed to consolidate COPE against the honest-but-curious and malicious attacks. Honest-but-curious attack serves adversaries who accurately obey the protocol but try to learn as much private information as possible for their curiosity. Additionally, this kind of attack is not destructive consequently. However, it may leave the backdoor for the more dangerous attacks carrying catastrophes to the system. Malicious attack tries to learn not only the private information but also modifies the packet on harmful purposes. In our work, the SCOPE protocol is defensive to the both attacks. The private information in the COPE packet are encrypted by Elliptic Curve Cryptography (ECC), and an additional information is inserted into SCOPE packets served for the authentication process using the lightweight hash Elliptic Curve Digital Signature Algorithm (ECDSA). We then prove our new protocol is still guaranteed to be a secure method of data coding, and to be light to effectively operate in the peer-to-peer wireless network.*

## 1. INTRODUCTION

To increase the network performance, network coding mechanisms are adopted to reduce as much bandwidth consumption as possible. The network coding can do it as the intersecting node (i.e., intermediate node) aggregates (i.e., codes) several packets into a coded packet, then it broadcasts the coded packet to all nodes in its radio range. In our work, we leveraged a simple but very effective network coding mechanism for wireless networks, that is, the Coding Opportunistically (COPE) model [1]. COPE is simply applied for any 2-hop paths in the wireless network. It uses a binary Exclusive-OR (XOR) to code several eligible packets into the coded packet. Moreover, not all packets can be coded at the intersecting node. Only eligible packets, satisfying the well-predefined coding conditions evaluated by the intersecting node, can be coded then broadcast to the intersecting node's neighbors.

Although COPE is simple and effective, it still owns a number of drawbacks leading to the security and privacy issues. The challenge, however, is to ensure the security and privacy of a network coding protocol without much sacrificing the initial goal of increasing network capacity. It is indeed well known that cryptographic schemes do mostly come with huge costs both in terms of size and processing time. Hence, we proposed a lightweight cryptographic approach, namely SCOPE, in securing COPE by adopting the Elliptic Curve Cryptography (ECC) algorithm [2]. SCOPE solves the security and privacy issues of COPE against the honest-but-curious attack. Honest-but-curious attackers accurately obey the protocol but try to obtain as much private information in the packets as possible. More particularly, SCOPE protects the packets against being overheard, so the packet content cannot be read easily by the other nodes. In another hand, the intermediate SCOPE nodes in charge of coding the packets cannot misuse the packet data. They cannot be aware of the packet queues of their neighbors, so cannot fetch packets that can be coded. SCOPE is still effective and efficient in encrypting and coding at the same time, as ECC is a lightweight encryption algorithm and most of its operation are binary operators such as AND, OR, Ex-OR (XOR), etc. Therefore, SCOPE fits the features of a peer-to-peer networks.

Even though SCOPE can protect the data coding against honest-but-curious attack, SCOPE still has short-coming in protecting the packets against malicious attack. Malicious attack is more dangerous than honest-but-curious attack, as it can modify or change the data contained in the packet. For example, let us see the network scenario in Figure 2. We can see that $N_1$ can listen to and get packets which $N_2$ and $N_4$ send to $N_5$. If $N_1$ is a malicious node, it can modify the packet's data from $N_2$ and $N_4$ for the other transactions, or modify the data. In this work, we exploit the authenticating strategy to make SCOPE more robust against the malicious adversaries by applying the Elliptic Curve Digital Signature Algorithm (ECDSA) [3]. Under the security power of ECDSA, the zombie packets are removed from the transaction and cannot reach the destination node.

The rest of this paper is organized as follows: in Section 2 we review the related works about secure data coding. In Section 3 we introduce COPE and the core ECC encryption algorithm. In Section 4 we formalize our security models. Section 5 discusses the security requirements necessary for the proposal. Section 6 introduces our proposal, namely SCOPE. In Section 7 we present the COPE and SCOPE packet format to make a clarification of the security property of the proposal. The lightweight cryptographic SCOPE against attacks are detailed in section 8. The experiments are evaluated in Section 9. The security properties are expressed in Section 10. We finally conclude the paper in Section 11.

## 2. RELATED WORKS

In the work [4], authors analyze the security of network coding mechanisms. This work considers the single source scenario in which an eavesdropper is able to read the private information. Subsequently, [5] secures network coding with different models has been proposed. In [6], authors propose a Wiretap model for collecting subsets of nodes in a network coding system such that each Wiretapper is allowed to access only one of these subsets. Moreover, we find another model proposed in [7] in which the authors focus on preventing traffic analysis and tracing in multi-hope wireless networks, by deploying Paillier [8], a homomorphic encryption scheme operating on large primes. Hence, the related performance as operations done on large prime numbers is quite costly. In [9], authors analyse a numerous shortcomings of the Homomorphic Message Authentication Code (H-MAC), that is, its vulnerability to pollution attacks, and how to secure communications in the transmission networks. The work is showed that the proposed method results in decreasing the related bandwidth overhead, but the overhead remains considerable. Additionally, a variety of the other related works [10, 11] also apply H-MAC based techniques to secure network coding; however, the overhead remains quite high and the problem of data and attack risk still remains an issue. Moreover, authors in [12] identify a flaw in H-MAC in general and have provided a corresponding inaccuracy in its formal security proof. This keeps it to doubt whether H-MAC based solutions are worthwhile in practice or not. Differently from the available works in cryptographic secure network coding schemes, our work adopts the lightweight ECC and ECDSA schemes to address the security and privacy issues. The work is applied into the COPE data coding scheme.

## 3. PRELIMINARY

In this section, the coding algorithm COPE and the core encryption algorithm Elliptic Curve Cryptography (ECC).

### 3.1 Coding Opportunistically (COPE)

COPE has been the first practical network coding mechanism significantly reducing a number of transmissions in the wireless networks, so well optimizes the wireless bandwidth consumption. A standard COPE is always applied for the 2-hop flows only. A 2-hop flow is considered as a path includes three nodes that are connected and involved in a transaction. Among those three nodes, the middle node is called an intersecting node. It does the function of an encoder, more specifically, it codes (i.e., aggregates) several packets into one packet. The outputted packet, namely coded packet, is then forwarded through a single transmission. Whereas, the two sided nodes are called respectively destination node (i.e., receiving the coded packet) and source node (i.e., sending the packet to the destination node through the intersecting node). The destination node is also considered as a decoder. It can decode the coded packet and obtain the packet to be claimed for it. Let us understand the above concepts through the following example.

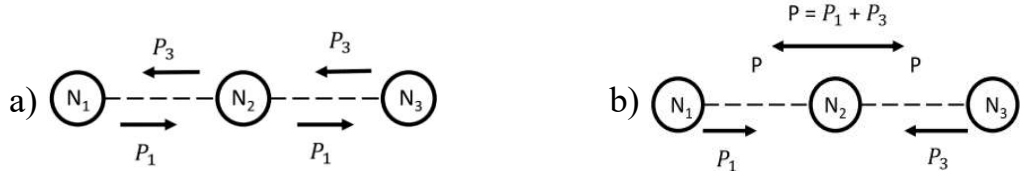

Figure 1: Simple data transmission model for three nodes. a) Without COPE; b) With COPE.

**Example 1.** Let consider 2-hop flows, that is, flow 1 $F_1 : N_1 \rightarrow N_2 \rightarrow N_3$ and flow 2 $F_2 : N_3 \rightarrow N_2 \rightarrow N_1$ in Figure 1. $N_2$ is called an intersecting node or encoder, since it locates between $N_1$ and $N_3$. In flow 1, node $N_1$ transfers the packet $P_1$ to $N_3$, thus $N_1$ is a source node and $N_3$ is a destination node or decoder. In flow 2, node $N_3$ transfers the packet $P_3$ to $N_1$, thus $N_3$ is a source node and $N_1$ is a destination node or decoder. Figure 1-a describes the network transmission without being applied with COPE. In that case, $N_2$ receives 2 packets $P_1$ and $P_3$ from $N_1$ and $N_3$, then creates 2 unicast transmissions to forward $P_1$ to $N_1$ and $N_3$, and 2 unicast transmissions to forward $P_3$ to $N_1$ and $N_3$. Whereas, Figure 1-b presents the network transmission applied with COPE. As $N_2$ receives $P_1$ and $P_3$ from $N_1$ and $N_3$ respectively, it codes $P_1$ and $P_3$ by Exclusive-ORing (i.e., XORing) them to produce a coded packet $P = P_1 + P_3$. Then, $N_2$ creates one broadcast transmission to forward P to $N_1$ and $N_3$. At $N_1$, $P_1$ decodes $P$ to achieve $P_3$ by XORing $P$ by its $P_1$, i.e. $P_3 = P + P_1$. The same decoding way is invoked at $N_3$.

### 3.2 Elliptic Curve Cryptography (ECC)

Let $P_1$ and $P_2$ be two considered messages, $(S_k, k)$ be secret key and public key, $P$ be the coded data of $P_1$ and $P_2$, $C$ be the cipher text of $P$ encrypted with $k$, $Deck(C)$ or $P$ be the plain text of $C$ decrypted with $k$. We have:

- Encryption: $C = Enc_k(P) = Enc_k(P_1 + P_2) = Enck(P_1) + Enck(P_2)$.

- Decryption: $P = Dec_{S_k}(Enc_{S_k}(C)) = Dec_{S_k}(Enc_{S_k}(P_1 + P_2)) = P_1 + P_2$.

In this work, we adopt Elliptic Curve Cryptography (ECC) based on the binary finite field $F_m^2$. Assume each node $n_i$ in the network has a pair of keys $(k_i, k_i.B)$ where $B$ is a base parameter of ECC and public to all over the network, $k_i$ is the private key, and $k_i.B$ is the public key. With ECC, the encryption of a message $P_i$ with $k_i$ is $C_i = Enc_{k_i}(P) = (r_i.B, P_i + r_i.B.k_i)$, and the decryption of $C_i$ with $k_i$ is $P_i = Dec_{k_i}(C_i) = P_i + r_i.B.k_i - (r_i.B).k_i$, where $r_i$ is a random number generated by the encryptor. Hence, the addition of two encryption gets $C = Enc_{k_i}(P_1) + Enc_{k_i}(P_2) = (r_1.B + r_2.B, P_1 + P_2 + r_1.B.k_i + r_2.B.k_i)$, and the responsive decryption is $Dec_{k_i}(C) = P_1 + P_2 + r_1.B.k_i + r_2.B.k_i - (r_1.B + r_2.B).k_i = P_1 + P_2$.

## 4. THREAT MODELS

In this paper, both of honest-but-curious attack and malicious attack are taken into a careful consideration and discussed, as they affect seriously to the systems. Observing them well can reduce the risks of leaking the private information of the system for any harmful purpose.

***Honest-but-curious attack.*** They correctly operate the protocol without modifying data. However, they try to learn as much personal information of the other users as possible to satisfy their curiosity. These adversaries do not cause serious consequences, but can leave a back door for the other attacks if they do not preserve well the learned data. In this work, each node may be considered as an honest-but-curious attacker. They can learn the private information from the COPE header as well as infer the path on which the packet moves through. If they are intersecting nodes, they can try on the received packets. In case they are surrounding nodes of the packet's path, they can try to overhear the packets from that path.

**Example 2.** The adversary can get the aggregate package, e.g., $P = P_1 + P_2 + P_3$, at the same time it receives another aggregate packet, i.e., $P' = P_1 + P_2$, so it can infer the content of the packet $P_3 = P - P'$.

***Malicious attack.*** It is much more hazardous than honest-but-curious attack. A malicious adversary tries to intervene as much user data as possible. They not only aim to learn as much personal information as possible, but also to modify data or replay it to achieve the further goals. Since this kind of attack can affect to the user personal information and the systematic accuracy, it is very challenging to be coped with.

**Example 3.** Let us continue Example 2. After inferring the content of the packet $P_3$, the adversary can even modify the COPE header of the packet as well as replace the packet $P_3$ with another packet $P_4$ by $P'' = P' + P_4$, then use $P''$ to forge $P'$, and send $P''$ to its destination.

## 5. SECURITY REQUIREMENTS

To avoid serious consequences of the above security threats, the security requirements need to be guaranteed on the aggregate packet moving through the network.

- ***Packet payload security.*** The packet payload cannot be read by any node in the network flow. The packet information should be accessed only by its claimed destination node. To satisfy this requirement, a solution should adopt cryptographic algorithms into the problem. The COPE packets should be encrypted with the public key of the destination node, and are aggregated in a secret way. Specifically, two parts need to be encrypted, that is, packet payload and fields in the COPE header for coding condition evaluation (see Section 7.1). The solution for this requirement will be presented in further details in Section 8.

- **Secure coding condition evaluation.** The intersecting nodes only code the packets they receive if the packets satisfy the coding conditions (see Definition 1, Section 8.1) set in every flows. However, this coding condition evaluation process possibly leaks the packet flow information. To avoid the other party to see the evaluation process at the intermediate node, the security solution is needed. In this work, the cryptographic solutions are proposed to secure this coding condition evaluation process. This will be expressed in more details in Section 8.

- **Packet data integrity evaluation.** When a packet arrives at a node in its flow, any intersecting node, receiving the packets to be coded, can modify the packet data during that coding process, or it can keep the packet replayed later on. In order to avoid this problem, a data integrity evaluation is needed. The one-way hash algorithm is adopted to make the authentication to be accurate. In this work, we adopt ECDSA as a valuable candidate for this problem. This solution will be presented in Section 8.

- **Cryptography-driven performance optimization.** Devices in the wireless network have many restrictions, i.e., they have limited physical resource and performance. Hence it is required to select a lightweight cryptographic algorithm to secure packets and make the protocol of encrypting packets more securely. In this work, the lightweight cryptographic algorithms, i.e., ECC and ECDSA, are adopted to reduce the performance cost, at the same time keep the security complexity of an encryption algorithm. The details will be expressed in Section 8.

# 6. SCOPE: A LIGHWEIGHT CRYPTOGRAGPHY SOLUTION

SCOPE is come up with the idea of how to combine the crucially common computing features of COPE and ECC, at the same time, to reserve objectives of COPE and secure the data. As mentioned above, COPE and ECC use the same binary operator that is XOR in coding or ciphering the data. COPE uses XOR to code different data (see Section 3.1) to output one coded data. Whereas, ECC is an additive homomorphic encryption so can aggregate different data into a coded data, and cover the coded data in the format of a cipher text (see Section 3.2). Hence, aggregating different pieces of data in secret way using ECC generates an encryption of a coded data. This is the core idea of SCOPE. Let us describe in more detailed of the SCOPE protocol.

Let us see Figure 2. The 2-hop flows of three nodes $N_i, N_m, N_j$ are considered in the both protocols. In the flow 1, node $N_i$ sends a packet $P_i$ to node $N_j$ through node $N_m$. Whereas, in the flow 2, node $N_j$ sends a packet $P_j$ sent from $N_i$ through $N_m$. Hence, $N_m$ is the intersecting node in either case. $N_m$ is in charge of checking the coding conditions on the received packets before coding the packets. The Figure 2-a depicts the COPE protocol without encryption. In that, $N_m$ invokes a XOR operator (denoted as "+") over the two received packets, and obtains an aggregate value, that is, $P_m = P_{ij} + P_{ji}$.

Whereas, the Figure 2-b depicts the SCOPE protocol with encryptions. Source node $N_i$ encrypts $P_{ij}$ by $N_j$'s public key $k_j$ and obtains an encryption $Enc_{K_j}(P_{ij})$. It then propagates this encryption to $N_m$. In the meantime, $N_j$ encrypts $P_{ji}$ by $N_i$'s public key $k_i$ and obtains an encryption $Enc_{k_i}(P_{ji})$. Then $N_j$ sends this encryption to $N_m$. The intersecting node $N_m$ aggregates the two received encryptions from $N_i$ and $N_j$ by performing the operator Exclusive-OR on the two cipher texts, to receive $C = Enc_{k_j}(P_{ij}) + Enc_{k_i}(P_{ji}) = Enc(k_i, k_j)(P_{ij} + P_{ji})$ where $C$ is the encryption of the coded data of $P_{ij}$ and $P_{ji}$. Then, $N_m$ broadcasts $C$ to both $N_i$ and $N_j$. With flow 1, the destination $N_j$ again adds its $Enc_{k_j}(P_{ij})$ to $C$ and obtains $Enc_{k_j}(P_{ij}) + (Enc_{k_j}(P_{ij}) + Enc_{k_i}(P_{ji})) = Enc_{k_i}(P_{ji})$. $N_j$ then decrypts $Enc_{k_i}(P_{ji})$ with its private key to achieve the data from $N_i$ claimed to it, that is, $P_{ji}$. The same steps are similarly performed at $N_i$ on the flow 2.

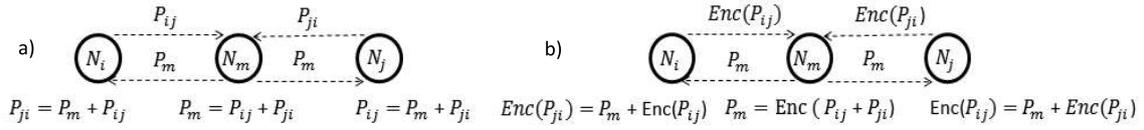

Figure 2: Protocols: COPE vs SCOPE - a) COPE; b) SCOPE.

## 7. Robust SCOPE Packet

In this section, COPE header format is presented. Then based on COPE header, a SCOPE header format is built up to empower COPE against the honest-but-curious and malicious attacks. More specifically, the key fields of COPE are encrypted using ECC with the public key of the destination node only which is allowed to read the packet data. Additionally, the COPE header has one more hashed field so that SCOPE can evaluate the integrity of the packet.

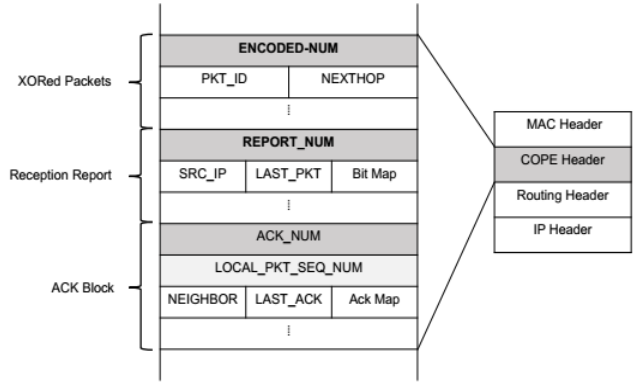

Figure 3: COPE Header Format.

### 7.1 COPE Header Format

A COPE header [1] (see Figure 3) is inserted into the header of a packet, placed between the MAC and IP headers. The COPE header has a structure, as follows. A COPE header includes three blocks, that is, coding report, reception report, and ACK report. Coding report contains information of the XOR-ed native packets and their next hop. Reception report contains information of overlearned packets from neighbors including the source, the received last packet from that source, and a bitmap presenting the list of recently received packets from that source. ACK report contains the information of received or missed packets which the sending node has, including a neighbor IP, the last ACKed packet from that neighbor and a bit map of ACKed packet.

## 7.2    SCOPE Packet

Based on the COPE header, in order to meet the security requirements mentioned in Section 5, the fields in the COPE header are encrypted with ECC in SCOPE, in terms of NEXTHOP, SRC_IP, LAST_PKT, Bit Map, NEIGHBOR, LAST_ACK, Ack Map, etc. Hence, all SCOPE nodes work on encryptions only whenever they access fields in the COPE header. SCOPE often access the header to read and compute the sets of neighbor nodes, next hops and previous hops, then used for evaluating the coding conditions. The coding conditions are defined in Definition 1, Section 8.1. For a reference to the coding condition evaluation, see Definition 3, Section 8.1.

| MAC Header | SCOPE Header | Routing Header | IP Header | Coded Packet |
|------------|--------------|----------------|-----------|--------------|

Figure 4: SCOPE Packet Format.

Moreover, to avoid the packets leaking the private information to COPE nodes, SCOPE encrypts the payload then sends it through the neighbors. If the packet arrives at an intersecting node, the intersecting node will code the received packets, then broadcast the outcome packet to the network.

## 7.3    Robust SCOPE Packet

SCOPE header as in Section 7.2 can protect the packet information against the honest-but-curious attack, since no node, except the destination node owning the public key used for encrypting the packet payload, can read the packet. Thus, the content of the packet is secret, and the honest-but-curious or malicious node even when it is an intersecting node still cannot read the packet content. An honest-but-curious adversary can release the packet after learning it, but a malicious adversary can harm the packet sequentially by dishonestly coding the encrypted packets with another fake packets just to break the protocol, or by modifying the packet payload.

Hence in this section, the SCOPE header is modified to be against malicious attack. Particularly, the payload of SCOPE is separated into the two parts, that is, the hash for packet authentication and the coded or encrypted payload.

| MAC Header | SCOPE Header | Routing Header | IP Header | SCOPE Header Signature | Coded Packet Signature | Coded Packet |
|------------|--------------|----------------|-----------|------------------------|------------------------|--------------|

Figure 5: Robust SCOPE Packet Format.

In the packet format, the hash of payload and the hash of SCOPE header both are generated using ECDSA [3]. The coded packet signature is made by the source node with its private key, and evaluated by the destination node with the public key of the source node. This signature is aimed to support the destination node to assess the packet if that packet is still original, or modified on the way to it. Hence, the integrity of packets can be guaranteed against the packet modification. Whereas, the signature of SCOPE header is made by the sending node with its private key. This signature is evaluated by the neighbor nodes using the public key of the sending node. The evaluation of this signature is aimed to support the receiving node in guaranteeing the received packets exactly from its neighbor.

# 8. Robust SCOPE Data Coding

## 8.1 Secure Coding Condition

Let us see Figure 2-b, node $N_i$ sends a packet $P_{ij}$ to its target, that is, $N_j$, through $N_m$. To make $P_{ij}$ able to reach $N_j$, the work at $N_m$ is crucial as it decides to forward the packet to $N_j$. Forwarding the packet is not simply receiving and sending the received packet $P_{ij}$ to the network. It does not also mean that packets move through the intersecting nodes to reach their destination, not all of intermediates will code the by-passed packets and propagate the result packets towards. As the objectives of coding protocol, to reduce the bandwidth cost, $N_m$ needs to aggregate (i.e., code) several packets together and forwards only the aggregate packet. In order to support $N_m$ in deciding the $P_{ij}$ propagation, the coding conditions are built up for $N_m$ to be evaluated. The necessary parameters for the coding condition evaluation are retrieved from the header of COPE packet. Only the packets satisfying the coding conditions will be aggregated into their suitable packet then sent towards to their destination. More specifically:

**Definition 1.** *Coding condition.* Let $F_i$ and $F_j$ be two flows of packets crossing at node $N_m$, i.e., $F_i \cap F_j = N_m$. $N_m$ codes packets received from $F_i$ and $F_j$ in case the next hops of the packet at $N_m$ on flow $F_i$ (or flow $F_j$) are the previous hops of the packet at node $N_m$ on flow $F_j$ (or flow $F_i$), or are the neighbors of that previous hop. More formally:

$$X_j = PH[N_m, F_j]$$
$$X_i = PH[N_m, F_i]$$

$$C(N_m, F_i, F_j) = \begin{cases} true, & ((NH[N_m, F_i] \subset NB_{X_j}) \vee (NH[N_m, F_i] = PH[N_m, F_j])) \\ & \wedge((NH[N_m, F_j] \subset NB_{X_i}) \vee (NH[N_m, F_j] = PH[N_m, F_i])) \\ false, & otherwise. \end{cases} \quad (1)$$

where

- $C(N_m, F_i, F_j)$ is the coding condition.

- $N_m$ is the intersecting node of two flows $F_i$ and $F_j$.

- $NH[N_m, F_i]$ is the set of next hops of node $N_m$ in flow $F_i$.

- $PH[N_m, F_i]$ or $X_i$ is the set of previous hops of node $N_m$ in flow $F_i$.

- $NB_{X_i}$ is the set of all neighbors of nodes in $X_i$ in flow $F_i$.

**Example 4.** Let us see Figure 1-b. It is noted that in case of COPE, the set of neighbor nodes and the set of previous nodes contain only one element. It is assumed that there are two flows of packets, that is, $F_1$ and $F_2$ ($F_1 : N_1 \rightarrow N_2 \rightarrow N_3$, and $F_2 : N_1 \rightarrow N_2 \rightarrow N_3$). $N_2$ is clearly the intersecting node of the two flows, so it is also the intermediate node where the coding can cause. Hence, the set of previous hops of $N_2$ on $F_1$ is $PH[N_2, F_1] = \{N_1\}$, whereas the set of next hops of $N_2$ on $F_1$ is $NH[N_2, F_1] = \{N_3\}$. In the meanwhile, the set of previous hops of $N_2$ on $F_2$ is $PH[N_2, F_2] = N_3$, whereas the set of next hops of $N_2$ on $F_2$ is $NH[N_2, F_2] = \{N_1\}$. For each node in sets of previous nodes of $N_2$ in $F_1$ and $F_2$, $N_1$'s neighbors is $NB_{N_1} = \{N_2\}$, and the set of $N_3$'s neighbors is $NB_{N_3} = \{N2\}$. Let us check the coding conditions in Definition 1, we see that the case $C(N_2, F_1, F_2) = true$ happens.

The above conditions are readable and stored in the header of each node. However, the coding condition evaluation process is done by the intersecting node (i.e., the intermediate node) $N_m$. Node $N_m$ also needs the information from its surrounding nodes, especially the nodes are the sender and the recipient of the packets through it, as in Figure 2-b, that is, $N_i$ and $N_j$. Hence, in case that $N_m$ is not an honest and benign node, this assessment process can leak the packet flow information to the intermediate, and cause a serious consequence to the data security and privacy as presented in Section 5. This process thus should be done in a secure way. In this work, we adopt the homomorphic encryption as an effective way to secure this process, particularly, ECC is used for securing data. More specifically, all information owned by $N_m$ are encrypted with a public key $K_m$. All data at $N_i$ and $N_j$ are respectively encrypted with the public keys of $N_i$ and $N_j$, that is, $K_i$ and $K_j$. It is noted that the atomic data which is encrypted is the ID of node's neighbors or previous hops or next hops. The encrypted atomic data is specified as in Definition 2.

**Definition 2.** *Atomic Data Cipher.* Let $K_2$ be the public key of node $N_2$. Let $D_2$ is the atomic data to be encrypted with $K_2$. Therefore, $Enc_{K_2}(D_2)$ is the encryption (i.e., cipher) of the atomic data $D_2$ encrypted with the key $K_2$ of node $N_2$.

**Example 5.** Let us continue Example 4. Let $K_m$ be the public key of $N_m$, $N_i$ be the previous hop of $N_m$ on flow $F_1$. Hence, $Enc_{K_m}(N_i)$ is the encryption of $N_i$'s ID. It is noted that $N_i$ is also considered as its own ID.

Each of nodes in the network keeps one previous hop list, one next hop list, and one neighbor list. Hence, for evaluating the coding condition, the intersecting node, i.e., $N_m$, exploits its previous hop list and next hop list, at the same time, requests each of its previous nodes for sending it their neighbor list. These lists are also the input parameters of the coding evaluation process. The coding condition parameters all are lists of atomic data ciphers, and defined in Definition 2. The lists of encryption defined in Definition 3 are also stored in the SCOPE header instead of the plain text as in COPE header.

**Definition 3.** *Coding Condition Parameter Cipher.* Let $N_t$, $F_i$ be respectively the considered node and the consider flow of data. Let $NH[N_t, F_i], PN[N_t, F_i], NBPN[N_t, F_i]$ respectively $N_t$'s previous hop list, $N_t$'s next hop list and the neighbor node lists of $N_t$'s previous nodes. From Definition 2 of atomic data cipher, for each element of the lists $NH[N_t, F_i], PN[N_t, F_i], NBPN[N_t, F_i]$, the respective cipher of the lists are denoted as $Enc_{K_t}(NH[N_i, F_i])), Enc_{K_t}(PN[N_t, F_i]), ENB_{PN}[N_t, F_i]$ and defined as follows:

$$Enc_{K_t}(NH[N_t, F_i]) = \{Enc_{K_t}(N_{(0,F_i)}), Enc_{K_t}(N_{(1,F_i)}), \cdots, Enc_{K_t}(N_{(n,F_i)})\}$$

$$Enc_{K_t}(PH[N_t, F_i]) = \{Enc_{K_t}(N_{(n+1,F_i)}), Enc_{K_t}(N_{(n+2,F_i)}), \cdots, Enc_{K_t}(N_{(n+m,F_i)})\}$$

$$ENB_{PH[N_t, F_i]} = Enc_{(K_{(n+1)}, K_{(n+2)}, \cdots, K_{(n+m)})}(NB_{\{N_{(n+1,F_i)}, N_{(n+2,F_i)}, \cdots, N_{(n+m,F_i)}\}})$$

$$= \bigcup_{u=n+1}^{n+m} Enc_{K_u}(NB_{N_{(u,F_i)}})$$

$$= \bigcup_{u=n+1}^{n+m} \{Enc_{K_u}(N_{(u,F_i,1)}), Enc_{K_u}(N_{(u,F_i,2)}), \cdots\}$$

**Example 6.** Let us continue Example 4 and 5. Let $N_2$, $K_2$ be respectively the considered node and its own public key. Let $F_1$ be the considered flow. $N_2$'s lists of next hops, previous hops, and its previous nodes' neighbor node lists on flow $F_1$ as follows: $Enc_{K_2}(NH[N_2, F_1]) = Enc_{K_2}(N(3,1))$; $Enc_{K_2}(PH[N_2, F_1]) = Enc_{K_2}(N_1)$; $Enc_{K_2}(NBPH[N_2, F_1]) = Enc_{K_2}(N_1), Enc_{K_2}(N_3)$ as $N_2$ on $F_1$ has two neighbors, that is, $N_1$ and $N_3$.

The coding condition evaluation is processed at the intersecting node but it needs a collaboration among multiple parties (i.e., the intersecting node, and its previous hop and next hop on the same flow) to support this evaluation process. For example, as in Figure 1-b, the coding condition evaluation is done by $N_m$ but it needs a collaboration among $N_i$, $N_m$ and $N_j$. However, as presented in Section 5, to prevent the risk of information violation, this collaboration needs to be processed secretly to avoid leaking a nod's private information to the others. In this situation, the information of $N_i$ and $N_j$ must be kept against the reading of $N_m$. Moreover, the nature of each coding condition evaluation is a comparison among elements of the two lists. Hence, to meet the security requirement in Section 5, this comparison is securely processed among encryptions of elements of the lists. For example, as in Definition 1, one of coding conditions is the comparison between the lists $NH[N_m, F_i]$ and $PH[N_m, F_j]$, whereas, the comparison between $NH[N_m, F_i]$ and $NBPH[N_m, F_i]$ is a series of comparisons between the list $NH[N_m, F_i]$ and each of lists in $NBPH[N_m, F_i]$ since $NBPH[N_m, F_i]$ contains many lists, each of lists contains the neighbor nodes relating to each of $N_m$'s previous nodes. As in Definition 3, the coding conditions contain the lists of encrypted elements. The comparison operators used for assessing the coding conditions are executed on the lists of encryptions. Thus, let us present one secure comparison between $NH[N_m, F_i]$ and $PH[N_m, F_j]$ done at $N_m$. The other comparisons in the coding conditions are similarly performed.

Let us consider the two original lists of $NH[N_m, F_i]$ and $PH[N_m, F_j]$ as follows:

$$NH[N_m, F_i] = \{N_{(0, F_i)}, N_{(1, F_i)}, \cdots, N_{(n, F_i)}\}$$
$$PH[N_m, F_j] = \{N_{(n+1, F_j)}, N_{(n+2, F_j)}, \cdots, N_{(n+m, F_j)}\}$$

First $N_i$ is in charge of generating the encryption of $NH[N_m, F_i]$ using its destination node's public key, that is $N_j$'s public key (i.e., $K_j$), to obtain $Enc_{K_i}(NH[N_m, F_i]) = \{Enc_{K_i}(N_{(0, F_i)}), Enc_{K_i}(N_{(1, F_i)}), \cdots, Enc_{K_i}(N_{(n, F_i)})\}$. In the meanwhile, $N_j$ is in charge of generating the encryption of $PH[N_m, F_j]$ with its destination node's public key, that is $N_i$'s public key (i.e., $K_i$), to obtain $Enc_{K_j}(PH[N_m, F_i]) = \{Enc_{K_j}(N_{(n+1, F_j)}), Enc_{K_j}(N_{(n+2, F_j)}), \cdots, Enc_{K_j}(N_{(n+m, F_j)})\}$. After generating the encryption lists, $N_i$ sends the encryptions to $N_m$, while $N_j$ sends the encryptions to $N_m$. Hence, $N_m$ can help transfer the received encryption lists to their destination nodes, that is, $N_m$ sends $Enc_{K_i}(NH[N_m, F_i])$ to $N_j$, and forwards $Enc_{K_j}(PH[N_m, F_i])$ to $N_i$. At $N_i$, $N_i$ continues to uses its public key, i.e., $K_i$, to encrypt the encryption list from $N_m$, to obtain a twice-encrypted list, that is, $Enc_{(K_i, K_j)}(PH[N_m, F_i]) = \{Enc_{(K_i, K_j)}(N_{(n+1, F_j)}), Enc_{(K_i, K_j)}(N_{(n+2, F_j)}), \cdots, Enc_{(K_i, K_j)}(N_{(n+m, F_j)})\}$. Similarly, $N_j$ again encrypts the received list of encryptions with its public key, i.e., $K_j$, and obtains the twice-encrypted list, that is, $Enc_{(K_j, K_i)}(NH[N_m, F_i]) = \{Enc_{(K_j, K_i)}(N_{(0, F_i)}), Enc_{(K_j, K_i)}(N_{(1, F_i)}), \cdots, Enc_{(K_j, K_i)}(N_{(n, F_i)})\}$. After that, $N_i$ and $N_j$ transfer the twice-encrypted lists to $N_m$. $N_m$ then performs the function $EqualList()$ as in Algorithm 1. $EqualList()$ evaluates if two lists are equal to each other. It inputs two lists, that is, $Enc_{(K_j, K_i)}(NH[N_m, F_i])$ and $Enc_{(K_i, K_j)}(PH[N_m, F_i])$, and returns a boolean result, that is, $'true'$ or $'false'$. $'true'$ is returned as the two encryption lists are equal, and $'false'$ as the two encryption lists are unequal.

More specifically, in the Algorithm 1, $N_m$ traverses the two lists $Enc_{(K_j, K_i)}(NH[N_m, F_i])$ and $Enc_{(K_i, K_j)}(PH[N_m, F_i])$ (lines 2,3) to evaluate the equality of elements of two lists by subtracting (or Exclusive-ORing) each element $\overline{x}$ of $Enc_{(K_i, K_j)}(PH[N_m, F_i])$ by each element $\overline{y}$ of

---

**Algorithm 1** *EqualList()*

---

**Input:**

$$Enc_{(K_i,K_j)}(NH[N_m,F_i]) = \{Enc_{(K_i,K_j)}(N_{(0,F_i)}), Enc_{(K_i,K_j)}(N_{(1,F_i)}), \cdots, Enc_{(K_i,K_j)}(N_{(n,F_i)})\}$$

$$Enc_{(K_i,K_j)}(PH[N_m,F_j]) = \{Enc_{(K_i,K_j)}(N_{(n+1,F_i)}), Enc_{(K_i,K_j)}(N_{(n+2,F_i)}), \cdots, Enc_{(K_i,K_j)}(N_{(n+m,F_i)})\}$$

**Output:** true | false

1: $count = 0$;
2: **for** $\overline{x} \in Enc_{(K_i,K_j)}(NH[N_m,F_i])$ **do**
3:     **for** $\overline{y} \in Enc_{(K_i,K_j)}(PH[N_m,F_j])$ **do**
4:         **if** $(\overline{x} + \overline{y} == Enc_{(K_i,K_j)}(0))$ **then**
5:             $count + +$;
6:         **end if**
7:     **end for**
8: **end for**
9: $sizeNH = sizeof(Enc_{(K_i,K_j)}(NH[N_m,F_i]))$;
10: $sizePH = sizeof(Enc_{(K_i,K_j)}(PH[N_m,F_j]))$;
11: **if** $(count == sizeNH == sizePH)$ **then**
12:     return true;
13: **end if**
14: return false;

---

$Enc_{(K_j,K_i)}(NH[N_m,F_i])$ (line 4). Let $count$ be a temporary integer. If the subtraction is equal to an encryption of 0 generated with the public key of $N_i$ and $N_j$, i.e., $K_i$ and $K_j$, that means $\overline{x}$ is equal to $\overline{y}$, $count$ is increased by 1 (line 5). Then, if $count$ is equal to the sizes of two lists, that is, $sizeNH$ and $sizePH$ (lines 9, 10, 11), the functions return $'true'$ (line 12), it means two lists are equal, otherwise false is returned (line 14). In case the two lists are equal, it also indicates that the coding condition is met. Sequentially, the other coding condition can be continued to be evaluated.

## 8.2 Secure Payload Coding

The fact that COPE header are protected against attacks of observing the flow of packet and intervening the packets' routines is protecting coding conditions and operations on them as presented in Section 8.1. Even though that is a meaningful and important security strategy, securing data payload also plays a substantial role since the payload contains several sensitive information of users. Especially, the coding is done at the intersecting node. As discussed in Section 4, the intersecting node can be an adversary, and the plain data can reveal the personal information to the intersecting node. Hence, the payload should be secured. In this work, ECC algorithm is used to encrypt data into the cipher. This solution makes the intersecting node unable to read the data but at the same time still work on the encryption only. Hence, the sending node needs to encrypt the data before propagating the encryption to the intersecting node.

**Definition 4. *Coded Payload Cipher.*** Let $K_0, K_1, \cdots, K_n$ be public keys of nodes $N_0, N_1, \cdots, N_n$. Let $Enc_{K_0}(P_0), Enc_{K_1}(P_1), \cdots, Enc_{K_n}(P_n)$ be the $n$ payload encryptions of packets: $P_0$ with $K_0$, $P_1$ with $K_1, \cdots, P_n$ with $K_n$. The coded payload cipher made at $N_m$ is formulated as follows:

$$Enc_{(K_0,K_1,\cdots,K_n)}P = Enc_{(K_0,K_1,\cdots,K_n)} \sum_{i=0}^{n}(P_i)$$

$$= \sum_{i=0}^{n}(Enc_{K_i}(P_i)) = \sum_{i=0}^{n}(r_i.B, P_i + r_i.B.K_i))$$

$$= (\sum_{i=0}^{n}(r_i.B), \sum_{i=0}^{n}(P_i + r_i.B.K_i))$$

where $r_0, r_1, \cdots, r_n$ are random numbers generated at nodes creating the partial encryptions.

**Example 7.** Let us continue Example 6. It is assumed that $N_i$ wants to send the packet $P_{ij}$ to $N_j$ through $N_m$ on flow $F_1$, so it encrypts the payload of $P_{ij}$ into $Enc_{K_j}(P_{ij})$ and forwards this encryption to $N_m$. In the meanwhile, $N_j$ wants to send the packet $P_{ji}$ to $N_i$ through $N_m$ on flow $F_2$, so it encrypts the payload of $P_{ji}$ into $Enc_{K_i}(P_{ji})$ and forwards this encryption to $N_m$. At node $N_m$, after evaluating the coding condition as in Example 4, $N_m$ code the two encryptions $Enc_{K_j}(P_{ij})$ and $Enc_{K_i}(P_{ji})$ by aggregating them, and get $Enc_{(K_i,K_j)}(P_{ij} + P_{ji}) = (r_i.B + r_j.B, (P_{ij} + P_{ji}) + (r_1.B.K_1 + r_2.B.K_2))$ as the coded payload cipher.

Receiving packets from different neighbor nodes, after evaluating the coding condition, $N_m$ detaches the encrypted payloads of all packets and code them. Then $N_m$ puts them into a new packet. Then, $N_m$ propagates it towards the network. As the receiving node gets the coded packet from $N_m$, it can assess the coded payload cipher, then decodes and decrypts the cipher to obtain the data. This process is concerned as the coded payload assessment. The decoded payload is defined as in Definition 5.

**Definition 5.** *Decoded Payload.* Let $N_n$ be the destination node receiving the encrypted coded payload as defined in Definition 4. Let $K_0, K_1, \cdots, K_{n-1}$ be public keys of $N_n$'s neighbor nodes $N_0, N_1, \cdots, N_{n-1}$. Let $Enc_{(K_0,K_1,\cdots,K_n)}\sum_{i=0}^{n}(P_i) = (\sum_{i=0}^{n}(r_i.B), \sum_{i=0}^{n}(P_i + r_i.B.K_i))$ be the coded payload cipher of packets from $N_0, N_1, \cdots, N_n$ with the random numbers $r_0, r_1, \cdots, r_n$ generated by nodes creating the partial encryptions. More formally, the decoded payload by $N_n$, that is $Enc_{K_n}(P_n)$, is defined as follows:

$$Enc_{K_n}(P_n) = Enc_{(K_0,K_1,\ldots,K_n)}\sum_{i=0}^{n}(P_i)) + Enc_{(K_0,K_1,\ldots,K_{(n-1)})}\sum_{i=0}^{(n-1)}(P_i))$$

$$= (\sum_{i=0}^{n}(r_i.B), \sum_{i=0}^{n}(P_i + r_i.B.K_i)) + (\sum_{i=0}^{(n-1)}(r_i.B), \sum_{i=0}^{(n-1)}(P_i + r_i.B.K_i))$$

$$= ((\sum_{i=0}^{n}(r_i.B) + (\sum_{i=0}^{(n-1)}(r_i.B)), (\sum_{i=0}^{n}(P_i + r_i.B.K_i) + \sum_{i=0}^{(n-1)}(P_i + r_i.B.K_i)))$$

$$= (r_n.B, P_n + r_n.B.K_n)$$

$$Dec_{K_n}(Enc_{K_n}(P_n)) = (r_n.B).K_n + P_n + r_n.B.K_n = P_n$$

**Example 8.** Let us continue Example 7. $N_j$ receives the coded payload cipher for it, that is, $Enc(K_i, K_j)(P_{ij} + P_{ji}) = (r_i.B + r_j.B, (P_{ij} + P_{ji}) + (r_i.B.K_i + r_j.B.K_j))$. $N_j$ still keeps the coded payload cipher of the packet it wants to send to $N_i$, that is, $Enc_{K_i}(P_{ji}) = (\eta_i.B, P_{ji} + \eta_i.B.K_i)$ where $\eta_i$ is a random number generated by $N_j$. Hence, the decoded payload is $Enc_{K_j}(P_{ij}) = Enc_{(K_i,K_j)}(P_{ij} + P_{ji}) + Enc_{K_i}(P_{ji}) = (\eta_i.B + \eta_j.B, (P_{ij} + P_{ji}) + (\eta_i.B.K_i + \eta_j.B.K_j)) + (\eta_i.B, P_{ji} + \eta_i.B.K_i) = ((\eta_i.B + \eta_j.B, ((P_{ij} + P_{ji} + (\eta_i.B.K_i + \eta_j.B.K_j + P_{ji} + \eta_i.B.K_i)) = (\eta_j.B, P_{ij} + \eta_j.B.K_j)$. Then, the decryption is executed at $N_j$ with the private key of $N_j$ (i.e., $K_j$), to get the data needed for $N_j$, i.e., $Dec_{K_j}(Enc_{K_j}(P_{ij})) = (\eta_j.B).K_j + P_{ij} + \eta_j.B.K_j = P_{ij}$.

## 8.3   Private Identification

In order to guarantee that the packet is not able to be modified by the malicious adversaries, in this work we do not focus on how to identify the impersonating nodes in the network. Instead, a secure protocol secretly analyzes and reasons if all changes made by a node on a packet are legal. To serve that goal, we ensure that the node can exactly authenticate its collaborative nodes that may be a contact or a source node. More particularly, *(1) Contact Signature* is used for the receiving node to evaluate a sending node to be exactly its neighbor. This avoids any node which does not meet the connecting conditions before making any transactions with the receiving node. In case the malicious nodes attempt to change the data of the SCOPE header while the packet is coming to the receiving node, the neighbor can be aware of that change based on the contact signature; *(2) Source Signature* is used for the destination node to check if the received packet is exactly received from the node claimed to be its source node, and the packet content has not been modified during the way to the destination node. This avoids the destination node to collect a fake or amended packet not claimed to be sent to it. Both signatures are generated by the ECDSA algorithm. ECDSA is a digital signature algorithm using Elliptic Curve Cryptography, and behind it, SHA-2 [13] is applied for creating the signature. To be more clearly, definitions and protocols of these identifications will be presented in detailed.

**Definition 6.  *Contact Signature.*** Let $K_n$ be the private key of the sending node. Let *SignEncode*, *SignReport*, *SignAck* be respectively the signature set of *Encoded_Num*, the signature set of *Report_Num*, and the signature set of *Ack_Num* in the COPE header format. The signature of SCOPE header, denoted as *SignSCOPE* at node $N_n$ is defined as follows:

$$SignSCOPE_{N_n} = \{\{SignEncode\}, \{SignReport\}\{SignAck\}\}$$
$$= \{\{\{(r_{E_0}, s_{E_0})_{K_n}, (r_{E_1}, s_{E_1})_{K_n}\}, \cdots\},$$
$$(r_{R_0}, s_{R_0})_{K_n}, (r_{R_1}, s_{R_1})_{K_n}, (r_{R_2}, s_{R_2})_{K_n} \cdots,$$
$$(r_{A_0}, s_{A_0})_{K_n}, (r_{A_1}, s_{A_1})_{K_n}, (r_{A_2}, s_{A_2})_{K_n}, (r_{A_3}, s_{A_3})_{K_n}, \cdots\}$$

where $E_i, R_i, A_i$ (with $i = 0, 1, \cdots$) are elements of the parts in SCOPE header; the pairs of $(r_x, s_x)$ are signatures of the SCOPE header's fields by using ECDSA with the field order in the part $x$.

It is reminded that the fields of SCOPE header are encryptions made by ECC as in Section 8.1. Hence, the hashed values in the SCOPE header signature are made from the encryptions of SCOPE header.

**Example 9.** It is assumed that $N_i$ sends a packet to $N_m$. The SCOPE signature of $N_i$ is generated with the private key of $N_i$ and achieve the SCOPE signature as follows: $SignSCOPE_{K_n} = \{\{\{(r_{PKT\_ID1}, sP_{KT\_ID1})_{K_n}, (r_{NEXTHOPE_1}, s_{NEXTHOPE_1})_{K_n}\}, \cdots\}, \{\{(r_{SRC\_IP_1}, s_{SRC\_IP_1})_{K_n},$ $(r_{LAST\_PKT_1}, s_{LAST\_PKT_1})_{K_n}, (r_{BITMAP_1}, s_{BITMAP_1})_{K_n}\}, \cdots\}, \{(r_{SEQ\_NUMBER}, s_{SEQ\_NUMBER})_{K_n})$
$\{\{(r_{NEIGHBOR_1}, s_{NEIGHBOR_1})_{K_n}), (r_{LAST\_ACK_1}, s_{LAST\_ACK_1})_{K_n}), (r_{ACK\_MAP_1}, s_{ACK\_MAP_1})_{K_n})\}, \cdots\}\}.$

---

**Algorithm 2** *evaluateContact()*

---

**Input:**

$$SignEncode = \{\{(r_{E_0}, s_{E_0})_{K_n}), (r_{E_1}, s_{E_1})_{K_n})\}, \cdots\}$$
$$SignReport = \{\{(r_{R_0}, s_{R_0})_{K_n}), (r_{R_1}, s_{R_1})_{K_n}), (r_{R_2}, s_{R_2})_{K_n})\}, \cdots\}$$
$$SignAck = \{(r_{A_0}, s_{A_0})_{K_n}), \{(r_{A_1}, s_{A_1})_{K_n}), (r_{A_2}, s_{A_2})_{K_n}), (r_{A_3}, s_{A_3})_{K_n})\}, \cdots\}$$
$$EncEncode = \{\{Enc_{K_d}(E_0), Enc_{K_d}(E_1)\}, \cdots\}$$
$$EncReport = \{\{Enc_{K_d}(R_0), Enc_{K_d}(R_1), Enc_{K_d}(R_2)\}, \cdots\}$$
$$EncAck = \{Enc_{K_d}(A_0), \{Enc_{K_d}(A_1), Enc_{K_d}(A_2), Enc_{K_d}(A_3)\}, \cdots\}$$

**Output:** true | false

1: **for** $(\overline{e} \in EncEncode\ AND\ \overline{f} \in SignEncode)$ **do**
2:     $HashEE = ECDSA(e);$
3:     **if** $(f! = HashEE)$ **then**
4:         return false;
5:     **end if**
6: **end for**
7: **for** $(\overline{r} \in EncReport\ AND\ \overline{s} \in SignReport)$ **do**
8:     $HashER = ECDSA(r);$
9:     **if** $(s! = HashER)$ **then**
10:         return false;
11:     **end if**
12: **end for**
13: **for** $(\overline{a} \in EncAck\ AND\ \overline{b} \in SignAck)$ **do**
14:     $HashEA = ECDSA(a);$
15:     **if** $(b! = HashEA)$ **then**
16:         return false;
17:     **end if**
18: **end for**
19: return true;

---

Since the SCOPE header is very essential in computing the key sets of next hops and previous hops which are subsequently used for evaluating the coding conditions at the intersecting nodes, node $N_i$ first has to sign the packet before it transfers a coded or native packet to its neighbors as an warranty of the packet's integrity. $N_i$ makes signatures for each value in the SCOPE header using ECDSA. After that, $N_i$ sends the signed packet to its neighbors. The receiving neighbor needs to evaluate if the integrity of the received packet is still withheld.

Algorithm 2 describes the procedure *evaluateContact*() in evaluating the SCOPE header signature. From the received packet, the node reads the SCOPE header to get the encryptions of necessary fields, stored in the set parameters *EncEncode*, *EncReport*, *EncAck*, at the same time, gets the hashed values from the SCOPE header signature, stored in the set parameters *SignEncode*, *SignReport*, *SignAck*. All are the inputted arguments of *evaluateContact*(). The function returns a Boolean value $'true'$ or $'false'$ to prove if the sending node is truly the neighbor of the receiving node. If the value $'true'$ returns, the sending node is the neighbor claimed to send the packet. Additionally, the packet is still not modified on the way to the receiving node. Algorithm 2 iteratively traverses the two sets *EncEncode* and *SignEncode* (line 1). For each element *e* of *EncEncode* and each element *f* of *SignEncode*, the Algorithm 2 hashes e using ECDSA with the receiving node public key and creates a signature of it, that is, $HashEE$ (line 2). Then, $HashEE$ is compared with $f$ (line 3). If they are inequivalent, the Algorithm 2 returns the value $'false'$ (line 4). Otherwise, the loop is continued until all elements in $EncEncode$ and $SignEncode$ are completely traversed. Consequently, the Algorithm 2 does the same steps with the two sets $EncReport$ and $SignReport$ (lines 7-12), then with the two sets, $EncAck$ and $SignAck$ (lines 13-18).

The *contact identification* aims to prevent any misprocess happening because the packet can be modified when moving to the receiving node. However, it cannot prove the integrity of the payload if any harmful changes come up. Thus, SCOPE is intensified in preventing the destination node to work on the masqueraded packet or the poisonous packet. In this work, the payload after encrypted is sequentially signed with the private key of the source node by using ECDSA. Hence, only the destination node can read the packet payload and authenticate with its private key.

**Definition 7. *Source Identification.*** Let $K_d$ be the private key of the source node used for generating the signatures. The signature of the payload is formulated as follows:

$$SignPayload_{K_d} = (r_{m_0}, s_{m_0})_{K_d}, (r_{m_1}, s_{m_1})_{K_d}, \cdots \qquad (2)$$

where $m_0, m_1, \cdots$ are the segments of the payload, the pairs of $(r_x, s_x)$ are the ECDSA signatures.

---
**Algorithm 3** *evaluatePayload()*

---
**Input:**

$$SignPayload = \{(r_{m_0}, s_{m_0})_{K_d}, (r_{m_1}, s_{m_1})_{K_d}, \cdots\}$$
$$CodedPacket = \{enc_0, enc_1, \cdots\}$$

**Output:** true | false
1: **for** ($\overline{e} \in CodedPacket\ AND\ \overline{s} \in SignPayload$) **do**
2:    $HashPM = ECDSA(e)$;
3:    **if** ($s! = HashPM$) **then**
4:       return false;
5:    **end if**
6: **end for**
7: return true;

---

Since the ECDSA has a restricted number of bytes to be the input data as it still uses a hash function underlying, say the SHA-2 algorithm, so the payload is divided into multiple data segments. All of them are iteratively hashed then generated the signatures based on those hashed values, and outputs a set of hash values. Algorithm 3 describes a function namely $evaluatePayload()$ that is executed for this purpose. The function return a Boolean value w.r.t. the input data. There are two input data, that is, the set of hash values of payload segments $SignPayload$, and the set of encryptions of payload segments $CodedPacket$. Algorithm 3 traverses the two sets, $SignPayload$ and $CodedPacket$ (line 1). For each element $e$ of $CodedPacket$, Algorithm 3 computes the hash value of $e$, that is, $HashPM$ using ECDSA (line 2). If the corresponding signature $s$ is unequal to $HashPM$, the functions stops and returns the Boolean value $'false'$. It implies that the payload verification fails. Otherwise, the loop continues until all signatures are evaluated to be true, then the function returns the Boolean value $'true'$.

## 9. EXPERIMENT

In this work, to prove for effectiveness and efficiency of the proposed secure protocol, experiments on the different amounts of nodes and different key sizes of ECC encryption are done. These

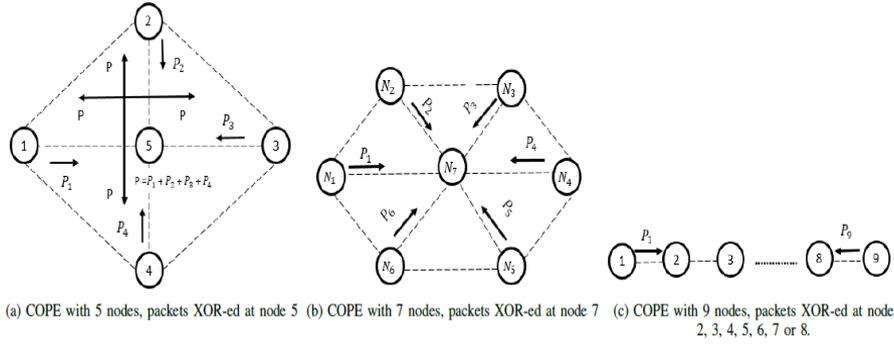

(a) COPE with 5 nodes, packets XOR-ed at node 5    (b) COPE with 7 nodes, packets XOR-ed at node 7    (c) COPE with 9 nodes, packets XOR-ed at node 2, 3, 4, 5, 6, 7 or 8.

Figure 6: SCOPE scenarios.

experiments are operated on a PC owning the physical resources in terms of CPU 1.8GHz Intel Duo-Core, RAM 4GB, HDD 16GB.

***Scenarios and Data.*** To assess the computing performance of SCOPE, we make a diversity of experiments on different parameters. Particularly, we create four SCOPE scenarios (as in Figures 1, 6-a, 6-b, 6-c), and change the key sizes of additive homomorphic ECC algorithm. Each calculated value in the experiments is the average of 20 times running the same experiments.

Table 1: Flows in test scenarios.

| Scenario | Feature | Flow |
|---|---|---|
| 1 | 1 | $F_1 : N_1 \rightarrow N_2 \rightarrow N_3; F2 : N_3 \rightarrow N_2 \rightarrow N_1$ |
| 2 | 6-a | $F_1 : N_1 \rightarrow N_5 \rightarrow N_3; F_2 : N_3 \rightarrow N_5 \rightarrow N_1;$ $F_3 : N_2 \rightarrow N_5 \rightarrow N_4; F_4 : N_4 \rightarrow N_5 \rightarrow N_2$ |
| 3 | 6-b | $F_1 : N_1 \rightarrow N_7 \rightarrow N_4; F_2 : N_4 \rightarrow N_7 \rightarrow N_1;$ $F_3 : N_2 \rightarrow N_7 \rightarrow N_5; F_4 : N_5 \rightarrow N_7 \rightarrow N_2;$ $F_5 : N_3 \rightarrow N_7 \rightarrow N_6; F_6 : N_6 \rightarrow N_7 \rightarrow N_3$ |
| 4 | 6-c | $F_1 : N_1 \rightarrow N_2 \rightarrow N_3 \rightarrow \cdots \rightarrow N_9;$ $F_2 : N_9 \rightarrow N_8 \rightarrow N_7 \rightarrow \cdots \rightarrow N_1$ |

Secure aggregation and evaluation time overload. Figures 1 and 6 describes four scenarios, and table 1 presents the number of flows w.r.t. the scenarios. Figure 1 involves 3 nodes and 2 flows. Figure 6-a involves 5 nodes and 4 flows. Figure 6-b involves 7 nodes and 6 flows. Figure 6-c involves 9 nodes and 2 flows. To evaluate the computing performance of SCOPE applied with ECC encryption algorithms, the selected ECC key sizes are varied in 163, 283, 409, 571 bits. These key sizes are guaranteed to be still secure by NIST.

First, the time on aggregating the payload ciphers in the four scenarios. These payload ciphers are aggregated using the additive property of homomorphic ECC algorithm. The number of payload cipher aggregation at the intersecting node in the four scenarios are respectively 2, 4, 6, 8 for each flow. In Figure 7-a, with the smallest ECC key size (i.e., 163 bits), the time on aggregating two payload ciphers of Scenario 1 is 26,8ms. In the worst case, that is, the largest key size (i.e., 571 bit), the time computed for aggregating 8 payload ciphers of Scenario 4 is 268,8ms. The times on different scenarios and key sizes are reasonable in the peer-to-peer environment.

Figure 7-b is another experiment to compute the time cost for SCOPE transmissions. These time are measured to evaluate the time which a packet moves through a flow from the source node to

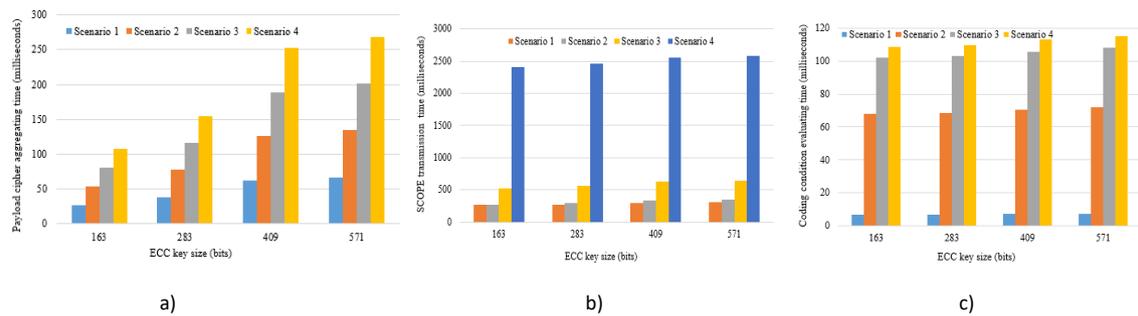

Figure 7: Time overload. *a) Time on aggregating the payload cipher at the intersecting node (milliseconds or ms) vs ECC key size (bits); b) Time on SCOPE transmission including aggregation time (milliseconds or ms) vs ECC key size (bits); c) Time on evaluating coding condition at intersecting node (milliseconds or ms) vs ECC key size (bits).*

the destination node. Hence, these times include the payload cipher aggregating times (as in the experiment of Figure 7-a) and transmission times. In this experiment, the number of payload cipher aggregations at the intersecting nodes are similar to the previous experiment. The ECC key sizes are also varied in 163, 283, 409, 571 bits. In the case of smallest ECC key size, that is, 163 bits, in scenario 1 involving 2 payload cipher aggregations, the time cost is 260,8ms. Whereas, in the case of largest ECC key size (i.e., 571 bits) and Scenario 4 involving 8 payload cipher aggregations, the time costs is 2.5s. The time overheads in this experiment in both cases are reasonable and prove that SCOPE is effective and efficient.

In order to evaluate the cryptographic process of evaluating the coding conditions of SCOPE, we measure the time SCOPE spent on this evaluation (Figure 7-c). The key sizes are selected in the range 163, 283, 409, 571 bits. The number of coding conditions for Scenarios 1, 2, 3, and 4 are respectively 4, 20, 30, 32 conditions. The time on evaluating the coding conditions in Scenario 1 (i.e., with the lowest number of conditions, that is, 4) with the smallest key size (i.e., 163 bits) is 6.8ms. In the meanwhile, the time on coding conditions evaluation in Scenario 4 (i.e., the highest number of condition, that is, 32) with the largest ECC key size (i.e., 571 bits) is 115,2 ms. In the both cases of the smallest parameters and the largest parameters, the time overheads are still reasonable and prove the effectiveness and efficiency of SCOPE.

***Private identification time overload.*** Moreover, for the private identification in the robust SCOPE, we made many experiments applying ECDSA on the packets. The selected ECDSA key sizes have 384 bits and 521 bits. First, the experiment on ECDSA are to consider how effective the ECDSA works (see Figure 8). We execute ECDSA to create a different number of signatures in the range of 5, 10, 15, 20. In the worst case with the maximum number of signatures (i.e., 20 signatures) and the largest ECDSA key size (i.e., 512 bits), the generation time is 1.8 seconds. In the best case with the minimum number of signatures (i.e., 8 signatures) and the smallest ECDSA key size (i.e., 384 bits), the generation time is 220ms.

The following experiments are done on combining ECC and ECDSA algorithms in the SCOPE protocol with the different number of signatures needed to be generated. In these experiments, ECC key sizes are in the range 163, 283, 409, 571 bits, ECDSA key sizes are in the range 384, 521 bits. The number of signatures need to be generated are in the range 5, 10, 15, 20. Figure 9-a presents experiments of encrypting pieces of data in the SCOPE header and payload, then generating signatures of their outputted encryptions. For each ECC key size and the 384-bit ECDSA key size, the time consumption is affected by the number of signatures too. In the worst case with the largest ECC key size (i.e., 571 bits) and 20 signatures, the time overload is 1.48 seconds. In the best case with the smallest ECC key size (i.e., 163 bits) and 5 signatures, the time overload is 270 milliseconds. Figure 9-b describes the time consumption for invoking the data encryption and generating the signature of the encryption using the 521-bit ECDSA key size. In the worst case with the largest ECC key size (i.e., 571 bits) and 20 signatures, the time overload is 2.4 seconds. In the best case with the smallest ECC key size (i.e., 163 bits) and 5 signatures, the time overload is 500 milliseconds.

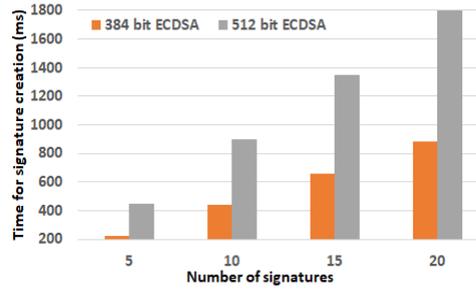

Figure 8: Signature Generation Time (ms) with different ECDSA key size bits.

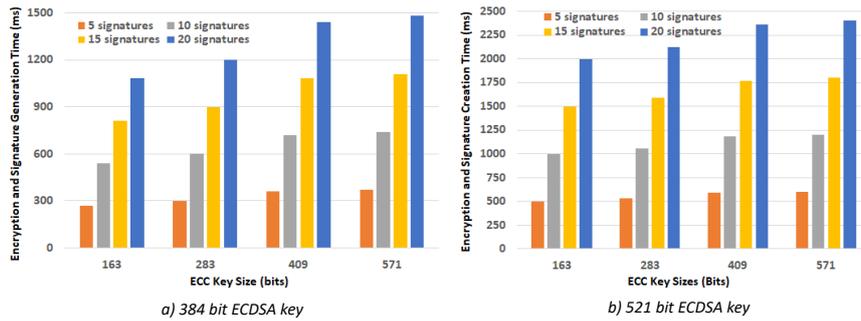

Figure 9: Time of executing data encryption by ECC and generating the signature of the encryption by ECDSA (ms).

## 10. SECURITY PROPERTY

In this section, a security proof is presented. More specifically, the expression how the proposal can cope with risks of honest-but-curious attack and malicious attack as in Section 4, as well as how the proposal meets the security requirements as in Section 5.

***Packet data security.*** The ECC homomorphic encryption is adopted to cipher the content of the data payload of packets. The payload then gets into a secret writing, i.e., unreadable. As a result, this carries the COPE packets a shield to deter adversaries from inferring any information inside the payload. Particularly, the ECC public key used for encrypting the payload is kept by only the destination node of the packet, so the other nodes cannot read the packet anyway. Only the destination node of the packet has the private key which can be used for decrypting the payload, then the payload can be read. Hence, both honest-but-curious and malicious adversaries cannot read the packet data.

***Information inferring protection.*** The addition property of the homomorphic encryption ECC is exploited for coding multiple packets into one packet at the intermediate node before the aggregate packet is propagated to the next hop of the intersecting node. More specifically, the public key of the destination node of the packet is used for this secret aggregation phase. The intersecting nodes just aggregates the encryptions without being aware of the packets' content. This point helps the data safe from the intersecting node. They cannot read the data inside and cannot infer the information as they do not have the private key of the destination node. This point thus help the data safe from both honest-but-curious and malicious adversaries.

***Coding condition evaluation security.*** The coding conditions involve encryptions of the node IDs as presented in Section 8.1. The comparisons are executed on these encryptions. Thus the intermediate nodes cannot learn the node IDs inside the thresholds and operands. Additionally, we also protect the neighbor nodes by encrypting their IDs, and only their direct neighbors can know their IDs, but the two-hop nodes cannot know their IDs. The comparative results are also not recognized by the nodes. Hence, the coding conditions are secured during the evaluation phase.

***Packet and node authentication.*** The digital signature algorithm based on Elliptic Curve Cryptography, namely ECDSA, is adopted in this work to intensify SCOPE against the malicious attack. This work focuses on authenticating the sending node and the original packet to avoid the malicious adversaries impersonating the sending node, or modifying the packet before reaching its destination node. For those purposes, the encryptions of SCOPE header fields are signed with its public key of the receiving node. So the receiving node can evaluate the SCOPE header fields' encryption before deciding to evaluate the coding conditions. This feature also helps SCOPE not let the packet go in case the packet meets the coding conditions and is coded then propagated to the destination node. Moreover, the source node in SCOPE makes the signature of the payload encryption with its private key of the source node. Hence, the destination node can use the public key of the source node to evaluate the signature to guarantee the integrity of the packet. It means that the packet payload is not changed by attackers on the channel from the source node to the destination node.

***Performance optimization.*** In this work, we empower our proposal's security with the ECC and ECDSA algorithms. The both algorithms are invoked based on the binary field with the binary operators. This point reduces much the time consumption, and meets the computing performance requirements. Additionally, the ECC is still guaranteed to be secure by NIST. So, the proposal can ensure the computing performance to be optimized.

## 11. CONCLUSION

In this work, we propose a cryptographic approach, namely SCOPE, which is able to support nodes secretly executing the COPE protocol against the honest-but-curious attack and the malicious attack, at the same time, consumes less time cost and get an effective computing performance due to the secure and lightweight ECC and ECDSA algorithms. The proposal is also proved to be effective and efficient through the different experiments on a variety of ECC and ECDSA key sizes for several scenarios. This work will be improved to fit with more network coding protocols, such as, BEND, DCAR, etc. in the future work.

## ACKNOWLEDGEMENT

This work is funded by the PSSG project at the Vietnamese-German University, Vietnam.